\documentclass[journal,final,letterpaper,twoside,twocolumn]{IEEEtran}
\usepackage[noadjust]{cite}      
\usepackage{graphicx}  
\usepackage{subfigure} 
\usepackage{amsmath}   
\usepackage{algorithm}
\usepackage{multirow}

\input{epsf}

\title{\Large\bf A Decision Feedback Based Scheme for Slepian-Wolf
Coding of sources with Hidden Markov Correlation}
\author{Krishna R. Narayanan and Kapil Bhattad
\\ krn,kbhattad@ee.tamu.edu
}
\begin{document}
\maketitle \pagestyle{empty}
\begin{abstract}
We consider the problem of compression of two memoryless binary
sources, the correlation between which is defined by a Hidden
Markov Model (HMM). We propose a Decision Feedback (DF) based
scheme which when used with low density parity check codes results
in compression close to the Slepian Wolf limits.
\end{abstract}

\thispagestyle{empty}


\section{INTRODUCTION}
Consider the classical Slepian Wolf set up where two correlated
sources $X$ and $Y$ have to be independently compressed and sent
to a destination. It was shown in \cite{SlepianWolf} that the
achievable rate region is $R_X \geq H(X|Y)$, $R_Y \geq H(Y|X)$ and
$R_X+R_Y \geq H(X,Y)$.  Recently, several practical coding schemes
have been designed for this problem based on the idea of using the
syndrome of a linear block code as the compressed output
\cite{pradhan}.  When $Y = X \oplus e$, where the sequence $e$ is
memoryless, low density parity check (LDPC) codes have been used
to achieve performance close to the Slepian-Wolf limit
\cite{Liveris}.

In this paper we consider the case when $Y = X \oplus e$, where
$X$ and $Y$ are binary i.i.d. sequences and $e$ is the output of a
Hidden Markov Model (HMM). This problem has been studied before by
Garcia-Frias {\em et al} \cite{Garcia1} and Tian {\em et al}
\cite{Garcia2}. In their scheme, $X$ is compressed to $H(X)$ bits
and transmitted. The encoder for $Y$ transmits a portion of the
source bits without compression to ``synchronize'' the HMM. The
remaining bits are used as bit nodes in an LDPC code and the
corresponding syndrome is transmitted. The decoder employs a
message passing algorithm with messages being passed between the
HMM nodes, the bit nodes and the check nodes. In \cite{Garcia2}
Tian {\em et al}, considered three HMM's and optimized the LDPC
code ensemble using density evolution for these specific models.
The resulting thresholds (the performance of an infinite length
LDPC code) were 0.08-0.12 bits away from the Slepian Wolf limits.

Here, we use a different approach. The main differences between
the proposed work and that in \cite{Garcia1,Garcia2} are that -
(i) a decision feedback scheme is used instead of iterating
between the HMM model nodes and the LDPC decoder. This also
reduces the decoding complexity significantly (ii) The LDPC codes
used are optimized for a memoryless channel instead of being
optimized for the channel with memory and, hence, the optimization
is considerably simpler than in \cite{Garcia2}. (iii) The proposed
scheme is similar to the scheme in \cite{Mushkin} to find the
capacity of the Gilbert-Elliott channel and is provably optimal
asymptotically in the length.

With the proposed scheme, for the models considered in
\cite{Garcia2} we are able to design codes that have thresholds
within 0.03 bits of the Slepian Wolf limits allowing for a
distortion of 1e-5, which is considerably better than those in
\cite{Garcia2}.

\section{PROPOSED SYSTEM}

\begin{figure}
\center{\includegraphics[scale=1]{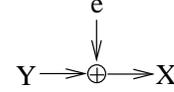}} \caption{Equivalent
Channel Coding Problem} \label{eqchnl}
\end{figure}

Consider two binary sources $X$ and $Y$ such that $Y = X \oplus e$
where $Y$ is independent and uniformly distributed. Typical
compression schemes to achieve a corner point in the Slepian Wolf
region involve sending $X$ using $H(X)$ bits and sending the
syndrome of $Y$ corresponding to a linear code ${\bf C}$ using
$H(Y|X)$ bits. It can be shown that the problem of compression is
equivalent to the problem of finding a capacity achieving linear
code for the channel shown in Fig. \ref{eqchnl} \cite{pradhan}.

When $e$ is memoryless, there are tools available to design LDPC
codes that achieve capacity on this channel and, hence, achieve
the Slepian-Wolf limit. In our case, $e$ is the output of a HMM
with three parameters $S$, $P$ and $\mu$. $S$ defines the
different states, $P$ is an $|S| \times |S|$ matrix with $P_{i,j}$
representing the probability of transition from state $S_{i}$ to
$S_{j}$ and $\mu$, $|S| \times 1$, has elements $\mu_{i}$ which
give $P(e=0|S_{i})$. The probability of $e$ being 0 or 1 depends
only on the current state. We further assume that when no state
information is available, the output of the HMM is equally likely
to be zero or one.

In \cite{Krishna} Narayanan {\em et al} use a Decision Feedback
Equalization (DFE) based scheme for ISI channels that makes the
channel appear memoryless to the LDPC decoder. We use the same
technique to make the channel appear memoryless and then design
codes for this ``memoryless'' channel.  The encoding and decoding
operations are explained below.

\subsection{Encoder}

\begin{figure}
\center{\includegraphics[scale=0.8]{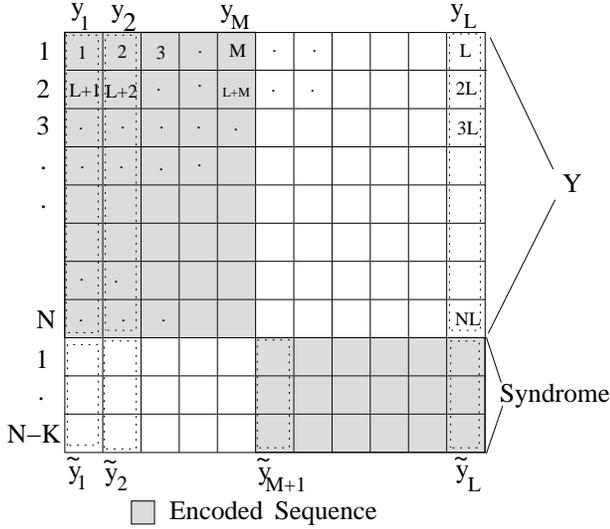}}
\caption{Encoded Sequence} \label{Encoder2}
\end{figure}

We will describe a scheme to achieve a corner point of the Slepian
Wolf coding region corresponding to $R_X = H(X)$ and $R_Y =
H(Y|X)$. The encoding process is shown in Fig. \ref{Encoder2}. Let
us assume that both sequences $X$ and $Y$ are first arranged in
the form of $L \times N$ matrices $\bf X$ and $\bf Y$. The
$(i,j)^{th}$ element in ${\bf Y}$ is $y_{(i-1)L+j}$. We will use
$Y_{i,j}$ to denote the $(i,j)^{th}$ element in $\bf Y$ and ${\bf
y}_j$ to denote the $j^{th}$ column of ${\bf Y}$. The sequence $X$
is compressed using an entropy coder to $H(X)$ bits. For the
models considered in this paper, the sequence $X$ contains
independent and uniformly distributed bits and, hence, no
compression is needed for $X$. The first $M$ columns in $\bf Y$
are transmitted without any compression (these are referred to as
pilots). For each column ${\bf y_j}, j > M$ in ${\bf Y}$ the
syndrome, ${\bf \tilde y_j}$, corresponding to an $(N, K)$ LDPC
code is computed and conveyed to the receiver. When an LDPC code
with $N$ bit nodes and $N-K$ check nodes is used, the syndrome
${\bf \tilde y_j}$ is simply the check values when the bit nodes
are set to ${\bf y_j}$. Therefore, the compressed sequence is
given by ${\bf Y_{comp}} = ({\bf y_1}, {\bf y_2}, \cdots, {\bf
y_M}, {\bf \tilde y_{M+1}}, \cdots, {\bf \tilde y_L})$. The
compression rate of this scheme is
\begin{equation}
R = \frac{NM+(N-K)(L-M)}{NL} = 1 - \frac{K}{N} \left(1 -
\frac{M}{L} \right)
\end{equation}

\thispagestyle{empty}

\subsection{Receiver}

The receiver has $\bf X$ and ${\bf Y_{comp}}$. Since the first $M$
columns in $\bf Y$ are sent without any compression, the receiver
has the first $M$ columns of $\bf Y$.  Hence, the receiver can
form the error values ${\bf e}_j$ for the first $M$ columns. From column $M+1$ onwards,
the receiver tries to recover ${\bf e}_j$
using the following procedure. It first computes soft estimates of
bits $e_{i,M+1}$ by using the error values in the past $M$
columns, i.e., $e_{i,1},e_{i,2},\ldots,e_{i,M}$ by using
\begin{equation}
\label{appLLR} \gamma_{i,M+1} = \log \frac {P(e_{i,M+1} = 1|
e_{i,M}, e_{i,M-1},\cdots, e_{i,1})} {P(e_{i,M+1} = 0| e_{i,M},
e_{i,M-1},\cdots, e_{i,1})}
\end{equation}
Note that the consecutive values of $e$ from any row are the
sequential outputs of the HMM and, hence, in Equation \ref{appLLR}
the estimate for a particular bit is made only from the past bits
in the same row. Since $e$ is the output of a HMM, $\gamma_{i,j}$
can be computed efficiently using the forward recursion of a BCJR
algorithm. From the soft estimates of $e_{i,M+1}$, one can
directly form soft estimates of $Y_{i,M+1}$ given by
$\lambda_{i,M+1}$ since $Y = X \oplus e$ and $X$ is available at
the receiver.

Now the LDPC decoder is run to decode ${\bf y}_{M+1}$ by using
${\bf \lambda}_{M+1}$ as the soft output corresponding to ${\bf
y}_{M+1}$ and ${\bf \tilde{y}}_{M+1}$ as the check values. With a
suitably chosen LDPC code the receiver can recover ${\bf
y_{M+1}}$. The whole process can be repeated to recover the next
column and so on till all columns are decoded. For an LDPC code
with finite length codewords, ${\bf y_{M+1}}$ will fail to decode
with some probability. This may cause error propagation within
that block.



\section{Achievable Information Rate}
The LDPC decoder tries to decode bits ${Y}_{i,j}$ by using
$\lambda_{i,j}$ which can be considered as the output of a channel
with input $Y_{i,j}$. If $L$ is made large the bits corresponding
to a particular column are far apart in time (at least $L$ time
units apart) and therefore it can be assumed that they go through
independent channels. That is, we can assume that for a given $j$,
the channel between $Y_{i,j} \rightarrow \lambda_{i,j}$ and
$Y_{p,j} \rightarrow \lambda_{p,j}$ are independent and identical
for $i \neq p$. The capacity of this channel is given by
\begin{eqnarray}
\nonumber
C & = & H(Y_{i,j} | X_{i,j}) - H(Y_{i,j} | X_{i,j},\lambda_{i,j})\\
\label{eqn:entropy}
& = & H(e_{i,j}) - H(e_{i,j} | \gamma_{i,j})
\end{eqnarray}
The second equality in Eqn. \ref{eqn:entropy} is true since $Y = X \oplus e$ and, hence,
$H(Y|X) = H(e)$. Since $\gamma_{i,j}$ is the optimal
estimate of $e_{i,j}$ given $e_{i,j-1}, \cdots, e_{i,j-M}$ we have
\begin{eqnarray}
\nonumber
C & = & H(e_{i,j}) - H(e_{i,j}|e_{i,j-1}, \cdots, e_{i,j-M}) \\
 &=& 1 - H(e_{i,j}|e_{i,j-1}, \cdots, e_{i,j-M})
\end{eqnarray}

If a capacity achieving code is used then the resulting
compression when $L \gg M$ is $H(e_{i,M+1}|e_{i,M}, \cdots,
e_{i,1})$. Note that the Slepian Wolf compression limit in this
case is $\lim_{M \rightarrow \infty} H(e_{i,M+1}|e_{i,M},
\cdots, e_{i,1})$. We can come arbitrarily close to the Slepian
Wolf limits by making $M$ large and using a capacity achieving
code for the ``memoryless'' channel. This shows the optimality of
this scheme for asymptotically large $L$ and $M$. Note that there
is a rate loss due to the first $M$ columns being transmitted
without compression, but that rate loss can be made arbitrarily
small by choosing a large enough $L$.

Although the arguments presented above show that this scheme is
optimal as $M \rightarrow \infty$, we do not require this. If we
use a code of rate $1-H(e_{i,j}|e_{i,j-1},\ldots,e_{i,1})$ for the
$j$th column, then we can obtain a compression rate of $\frac 1 L \sum_{j}
H(e_{i,j}|e_{i,j-1},\ldots,e_{i,1})$ which converges to $H(e) =
H(Y|X)$ from above as $L \rightarrow \infty$ for any wide sense
stationary process $e$. This solution however requires variable
rate LDPC codes for the different columns and, hence, is not used
in this paper.

\section{Simulation Results}
We compare the performance of the proposed scheme with the scheme
used in \cite {Garcia2}. The HMM used in \cite {Garcia2} has two
states $S_0$ and $S_1$ and is defined by four
parameters $P(S_0 \rightarrow S_0), P(S_1 \rightarrow S_1),
P(0|S_0), P(1|S_1)$.
The models considered are \\
M1: (0.01, 0.065, 0.95, 0.925) \\
M2: (0.97, 0.967, 0.93, 0.973) \\
M3: (0.99, 0.989, 0.945, 0.9895)\\
Note that the parameters in the model are chosen so that they
satisfy $P(e=0) = 0.5$.

\begin{figure}
\center{\includegraphics[scale=.5]{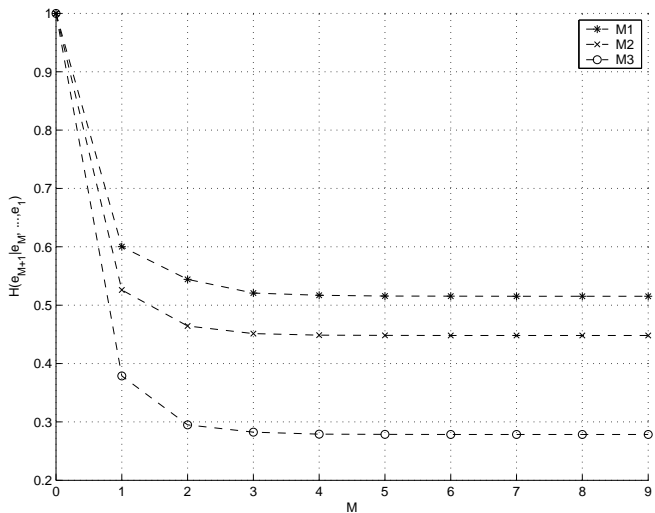}}
\caption{Compression Limit vs. $M$} \label{entropy}
\end{figure}

In Figure \ref{entropy}, we plot $H(e_{M+1}|e_{M}, \cdots, e_{1})$
as a function of $M$ for the  models. We observe that for these
models the $M$ required to come close to the Slepian Wolf
limits is quite small. We use $M = 4$ for our simulations.

For the three models the pdf of $\gamma_{i,j}$ conditioned on
$e_{i,j}$ being a 1 and 0 were computed through monte-carlo
simulations. From this the distribution of $\lambda_{i,j}$
conditioned on $y_{i,j}$ can be computed. Using this, an
 LDPC code ensemble was designed using density
evolution and differential evolution. It was assumed that a
Hamming distortion of $10^{-5}$ is acceptable. Since the HMM is
not symmetric, the pdf of $\gamma_{i,j}$ conditioned on $e_{i,j}$
being a 1 or 0 is not symmetric. That is, $f_{\gamma_{i,j}}(x |
e_{i,j}=1) \neq f_{\gamma_{i,j}}(-x| e_{i,j}=0)$, for some $x$.
Hence the distribution of $\lambda_{i,j}$ is also not symmetric.
For the density evolution we use the average of these pdf's
similar to the approach in \cite{wang}, where the correctness of
this procedure is proved. Simulations were done with the designed
LDPC codes of  length 100000 and $L=100$. 2000 such blocks were simulated for each model.
 A different interleaver was used
for each column to avoid repetition of error sequences. The
results obtained are compared with those of \cite{Garcia2} in
Table \ref{resTable1}. The SW limit column shows the Slepian Wolf
compression limit. The THEO column represents the threshold, which
is the achievable compression rate with infinite length LDPC codes.

For the DFE scheme simulations were also performed with $N =
2000$, $L = 100$ and $M = 4$. Codes designed for AWGN channel were
used in these simulations. The bit filling algorithm
\cite{Campello} was used to reduce error floors. The results are
also tabulated in Table \ref{resTable1}. For each model 5000
blocks were simulated. The Hamming distortion observed was less
than 2e-7. Although the performance in this case seems to be far
from the Slepian Wolf limits, it should be noted that this scheme is
universal and does not require any optimizations specific to the
HMM. Although beyond the scope of this paper, we wish to point out
that for small $L$ and finite lengths, simple improvements to the
decoding algorithm can provide significant improvements in the
compression rates. For example, allowing for decoding of a
particular block using the pilots on both sides.

The loss in rate due to the pilots in
the DF Scheme is not included in Table  \ref{resTable1}. If the pilots
are sent without compression, then the compression rate would increase
by 0.04. However, this loss can be reduced significantly by increasing $L$
and by compressing the pilots.

\begin{table}[h]
\caption{Results}
\begin{center}
\label{resTable1}
\begin{tabular}{|c|c|c|c|c|c|}
\hline \multirow{2}{*}{Model} & SW Limit & Tian {\em et al}\cite{Garcia2}
& \multicolumn{3}{c}{DF Scheme} \\
&$H(Y|X)$& THEO & THEO & $N=10^5$ & $N = 2000$\\
\hline
1 & 0.515 & 0.599 & 0.546 & 0.58 & 0.69\\
\hline
2 & 0.448 & 0.544 & 0.476 & 0.52 & 0.62\\
\hline
3 & 0.278 & 0.413 & 0.305  & 0.34 & 0.45\\
\hline
\end{tabular}
\end{center}
\end{table}

With $L = 100$ error propagation is a serious problem but it can
be overcome by lowering the rate of the LDPC code. In our
simulations, no error propagation was observed.


\thispagestyle{empty}

\section{Conclusion}
We proposed a low complexity decision feedback based scheme to
compress multiterminal sources with hidden Markov correlations.
The proposed scheme has thresholds just 0.03 bits away from the
Slepian Wolf limits and the simulated performance with designed
LDPC codes of length 100000 is within 0.08 bits of the limits
which is better than the thresholds of the scheme in
\cite{Garcia2}.

\bibliographystyle{IEEEtran}

\end{document}